# Sparse Signal Reconstruction with QUBO Formulation in $l_0$-regularized Linear Regression

Naoki Ide
*Advanced Research Laboratory, R&D Center, Sony Corporation*
Tokyo, Japan
Naoki.Ide@sony.com

Masayuki Ohzeki
*Sigma-i Co., Ltd, Tohoku University, Tokyo Tech*
Tokyo, Japan
mohzeki@sigmailab.com

*Abstract*— An $l_0$-regularized linear regression for a sparse signal reconstruction is implemented based on the quadratic unconstrained binary optimization (QUBO) formulation. In this method, the sparse signals are quantized and expressed as bit sequences. By transforming $l_0$-norm to a quadratic form of these bits, the quadratic objective function is provided to be optimized by the solver specialized for QUBO, such as the quantum annealer. Numerical experiments with a commercially available quantum annealer show that the proposed method demonstrates the reconstruction success rate nearly equivalent to conventional methods based on orthogonal matching pursuit (OMP) and least absolute shrinkage and selection operator (LASSO) under several conditions.

## I. INTRODUCTION

Sparse signal reconstruction is a technique that recovers the original signals from relatively few observed signals. Although this problem is usually ill-posed, there are several algorithms that perfectly recover the original signals with sufficient sparsity [1,2].

Direction-of-arrival (DOA) estimation in array signal processing is one of the applications of the sparse signal reconstruction. In this application, the directions of the signal sources such as sound sources and radar targets are estimated through the use of the sparse signal reconstruction schemes[3]. Since the signal sources usually appear sparsely along the azimuth axis, the distribution of these sources can be estimated through the use of such schemes.

In these sparse signal reconstruction techniques, the algorithms are usually based on $l_1$-regularized linear regression (LASSO) [1], which is an unconstrained extension of $l_1$-norm minimization that minimizes the $l_1$-norm, the sum of absolute values of the signal vector. When the order of the regularization is generalized as $p$, the objective function of the $l_p$-regularized linear regression is described as,

$$L_p(z) = \frac{1}{2\gamma_p} \|x - Az\|_2^2 + \|z\|_p, \quad (1)$$

where $z \in \mathbb{C}^M$ is a vector for the sparse representation, $x \in \mathbb{C}^N$ is a vector for the observed signals, $A \in \mathbb{C}^{N \times M}$ with ($N < M$) is an observation matrix, $\gamma_p$ is a regularization parameter and $\|z\|_p$ represents $p$-norm of $z$. For $p = 1$, $\|z\|_1$ is an $l_1$-norm of the vector to be minimized.

In $l_0$-norm minimization, an $l_0$-norm, the number of nonzero entries in sparse signal, is concern. Several theoretical considerations indicate that the $l_0$-norm minimizations are in principle superior to the $l_1$-norm minimizations [4]. However the implementation is practically difficult due to the potential threat of the combinatorial explosion, which forces us to use approximate approaches based on greedy policy, such as orthogonal matching pursuit (OMP) [5].

Quantum annealing [6] is a method that can tackle such laborious computations. The method efficiently solves a certain type of optimization problems using a QUBO formulation. The objective function of QUBO is expressed as a quadratic form of $n$ binary variables:

$$L(b) = b^T Q b, \quad (2)$$

where $b \in \{0,1\}^n$ is a vector of binary variables, and $Q \in \mathbb{R}^{n \times n}$ is a matrix of quadratic coefficients.

In this paper, we propose a method for the sparse signal reconstruction that solves the $l_0$-regularized linear regressions with quantum annealing. In the following sections, we explain the QUBO-based technique, and compare the performance of the proposed method with that of the conventional methods.

## II. SPARSE SIGNAL RECONSTRUCTION

### A. Conventional techniques

To solve a QUBO by quantum annealing, the binary variables to be optimized need to be expressed as variables $s \in \{-1,1\}^n$. When $b = (1-s)/2$ is substituted into (2), the objective function for the quantum annealing is expressed as

$$H(s) = -\sum_{i=1}^{n}\sum_{j<i}^{n} J_{ij}s_i s_j - \sum_{k=1}^{n} h_k s_k, \quad (3)$$

where $J \in \mathbb{R}^{n \times n}$ and $h \in \mathbb{R}^n$ are the coefficient matrices. When the coefficients $\{J, h\}$ in $H(s)$ are given to the quantum annealer the solution $s^*$ is provided and the solution $b^*$ is obtained by $b^* = (1-s^*)/2$.

The function $H(s)$ is known as the energy function for the spin system of the Ising model. Thus, $H(s)$ is minimized when the spin system achieves the ground state of the model. To achieve the ground state effectively, the quantum annealer uses an annealing process that controls quantum fluctuation appropriately to explore the solution space. The annealing process is performed in the discrete time steps. In each step, sufficient time is given in order for the system to converge to the ground state of the following Hamiltonian:

$$H_t(\sigma) = (1-\Gamma_t)H_p(\sigma^z) - \Gamma_t \sum_{k=1}^{n} \sigma_k^x, \quad (4)$$

where $\sigma$ is the Pauli spin operators for $n$ spins, and $\sigma^z, \sigma^x$ are $z$ and $x$ elements of $\sigma$, respectively. In (4), the second term represents quantum fluctuation, and $\Gamma_t$ is a parameter that controls the balance between two terms at step $t$. While the first term provides the system with the convergence to the ground state (QUBO solution), the second term provides the system with the superposition of all possible vectors to explore the solution space. Typically, $\Gamma_t$ starts from 1 for the initial superposition, and it gradually decreases to 0 to converge the system to the ground state. At the last step, the annealer provides a solution $s^*$ as the measurement value of $\sigma^z$.

The mechanism of quantum annealing is similar to simulated annealing, which numerically simulates thermal fluctuation. However, their performances are different because the quantum fluctuation generating tunneling effect, which enables state transition over a relatively large energy barrier in (3). Theoretical analysis has proved the superiority of the quantum annealing on the computation time over the simulated annealing [7]. However, the proof was not the theoretical guarantees for drastic acceleration of computation time like factorization in quantum gate computing.

The quantum annealing hardware system produced by D-Wave Systems [8] demonstrates quantum annealing at the device level. In this system, the variables $s$ and the coefficients $\{J, h\}$ are embedded on the qubit devices and their hardware parameters. The number of the available qubits (spins) is approximately 5000. In the implementation of the spin system, the physical network of qubits devices is automatically generated with the program codes using the dedicated software libraries. In addition, there is also software tool to extract the input parameters $\{J, h\}$ from the program codes about the objective function instead of an explicit expression of $Q$ [9]. Therefore, research topics on quantum annealing have spread to the QUBO formulation methods for various combinatorial optimization problems. For example, the Karp's NP-complete problems are formulated as QUBO in [10]. Several studies on robotics and communications [11,12] have provided QUBO formulations for the methods that were hampered by combinatorial explosions, and showed the potential of the methods using quantum annealing.

In [13,14], binary compressed sensing is formulated by QUBO and implemented using the quantum annealer. Binarizing the sparse vector is a simple approach to linearizing the nonlinear regularization term. However, this approach can result in inadequate resolution. Also, since there is no difference between $l_1$-norm and $l_0$-norm for the binarized $z$, it is uncertain whether the benefits of $l_0$-norm minimization will be obtained. To avoid the problem, the multilevel quantization for the sparse representation is used in the proposed method.

### B. Proposed technique

We propose a $K$-bit-quantized technique ($K > 1$), which decomposes a representation vector $z \in \mathbb{R}^M$ as

$$z = Bw, \quad (5)$$

where $B \in \{0,1\}^{M \times K}$ is a matrix of binary variables to be optimized, and $w \in \mathbb{R}^K$ is a weight vector. If $w_k (k = 1 \dots K)$ is $2^{-k}$, the quantization covers the representation value in range [0,1] with $2^K$ grids. If $w_1$ is $-2^{-1}$, the range moves to [-0.5, 0.5] and both positive and negative values are available. This quantization provides the following expression of the objective function for the $l_0$-regularized linear regression:

$$L(B) = \frac{1}{2\gamma_0}\|x - ABw\|_2^2 + \|Bw\|_0, \quad (6)$$

where $\gamma_0$ is a regularization coefficient.

To utilize the quantum annealing processing, each term of (6) has to be quadratic. While the first term is obviously quadratic, the second term is not apparent. To express the second term in a quadratic form, we formulate the second term with binary variables, and transform it into a quadratic form.

When a number $z$ is expressed by a bit sequence as $z = (b_1, b_2 \dots b_K)$, the product of $1 - b_k$ ($k = 1 \dots K$) takes 1 for $z=0 =(0,0\dots0)$ and otherwise takes 0 for nonzero $z$. Thus, the $l_0$-norm (number of nonzero elements) in (6) is expressed as:

$$\|Bw\|_0 = \sum_{i=1}^{M}\left(1 - \prod_{k=1}^{K}(1 - b_{ik})\right), \quad (7)$$

where $b_{ik}$ is the $k$-th bit for the $i$-th representation entry. The value inside the large parenthesis takes 1 if any of $b_{ik}(k = 1 \dots K)$ takes 1, which means $z_i$ is nonzero.

Although (7) is a binary form, it is not quadratic for $2 < K$ and needs to reduce the dimension for the QUBO formulation. For this reduction, it is effective to replace products of two binary variables with another binary variable, because the product also takes 0 or 1. To reduce the dimension of (7), the product in (7) is replaced with auxiliary variables $C \in \{0,1\}^{M \times (K-2)}$ as following way:

$$\prod_{k=1}^{K}(1 - b_{ik}) = c_{i1}(1 - b_{i3}) \dots (1 - b_{iK})$$
$$= c_{iK-2}(1 - b_{iK}), \quad (8)$$

where $c_{ik}(k = 1 \dots K - 2)$ is the $i, k$ element of $C$.

Although (8) is quadratic, the optimization of (8) is different from QUBO, because it requires the following constraints for the variables $c_{ik-1}, c_{ik}$, and $b_{ik+1}$,

$$c_{ik} = c_{ik-1}(1 - b_{ik+1}), \forall k = 1,2,\dots,K-2, \quad (9)$$

with $c_{i0} = 1 - b_{i1}$.

To formulate a constraint in an unconstrained optimization, it is usual to add a penalty term, which takes 0 only when the constraints are satisfied otherwise takes a positive value. When a constraint $c = ab$ is required for binary variables $a, b$, and $c$, a possible scheme that provides appropriate penalty is $(c - ab)^2 = c + ab - 2abc$. However, it still contains a triple factor term, $abc$. To remove $abc$, we further add $2(1-a)(1-b)c$, which also takes 0 when $c = ab$. Then, the final penalty term is

$$p(a, b, c) = 3c + ab - 2ac - 2bc, \quad (10)$$

With Equation (8), (10) and a positive real parameter $\lambda_c$, the objective function is now described:

$$L_p(B, C) = \frac{1}{2\gamma_0}\|x - ABw\|_2^2$$
$$+ \sum_{i=1}^{M}\left(1 - c_{iK-2}(1 - b_{iK})\right.$$
$$\left.+ \sum_{k=1}^{K-2}\lambda_c p(1 - b_{ik+1}, c_{ik-1}, c_{ik})\right). \quad (11)$$

The objective function obtained here becomes fully quadratic in terms of $B$ and $C$. With the parameters $\{J, h\}$ available from the observations, (11) can be evaluated using the quantum annealer.

## C. Extension for multiple observation

In this case, the reconstruction system is assumed to use the multiple observations from the different sparse vectors. These vectors are assumed to be duplicated from an original sparse vector with some fluctuation in the nonzero values.

The $l_1$-SVD (singular value decomposition) [3] is a method that handles these kind of the observations by group $l_1$-regularized linear regression (group LASSO). When the $S$ observations are obtained as $X = (x_1 \dots x_S) \in \mathbb{C}^{N \times S}$, the objective function of the group-LASSO is expressed as:

$$L(Z) = \frac{1}{2\gamma_1} \|X - AZ\|_2^2 + \|l_2(Z)\|_1, \quad (12)$$

where $Z = (z_1 \dots z_S) \in \mathbb{C}^{M \times S}$ is a matrix for the sparse vectors, and $l_2(Z)$ is a vector function that calculates the $l_2$ norms for $Z$ along the axis for $S$.

In addition, this method reduces the computational costs with an SVD-based preprocess. The proccess replace $X$ with a matrix $X_L = (u_1 \dots u_L) \in \mathbb{C}^{N \times L}$ from the decomposed form $X = U\Sigma V$. The substitution causes the sparse signals different from $Z = (z_1 \dots z_S) \in \mathbb{C}^{M \times S}$. However, it can be proved that the nonzero positions in $Z$ do not change by this modification.

To accommodate the scenario for the multiple observation, we modify (5) to the following form:

$$Z = Bw, \quad (13)$$

where $B \in \{0,1\}^{M \times L \times K}$ is the $M \times L \times K$ tensor with binary variables to be optimized, $L$ is the number of the observations and $w \in \mathbb{R}^K$ is the weight vector as same in (5). The matrix $Z \in \mathbb{C}^{M \times L}$ consists of the sparse vectors $z_1, z_2 \dots z_L \in \mathbb{C}^M$ whose nonzero positions are also the nonzero position in the original sparse vector $\zeta \in \mathbb{C}^M$. From this regulation, the number of the nonzero values in $\zeta$ is inferred greater or equal to the number of the nonzero rows in $Z$, which allows approximation of $\|\zeta\|_0$ with $\|l_2(Z)\|_0$. Thus, the objective function is expressed as

$$L(B) = \frac{1}{2\gamma_0} \|X - ABw\|_2^2 + \|l_2(Bw)\|_0. \quad (14)$$

Similar to (8), the $l_0$-norm group can be obtained as:

$$\|l_2(Z)\|_0 = \sum_{i=1}^{M} \left(1 - \prod_{l=1}^{L}(1 - c_{ilK-1})\right), \quad (15)$$

where $C \in \{0,1\}^{M \times L \times (K-1)}$ is the auxiliary binary variables. Similar to the derivation for (11), we obtain the objective function of the $l_0$-norm-group-regularized linear regression as

$$L_p(B,C,D) = \frac{1}{2\gamma_0} \|X - ABw\|_2^2$$
$$+ \sum_{i=1}^{M} \left(1 - d_{iL-1}\right.$$
$$+ \sum_{l=1}^{L-1} \left(\lambda_c p(1 - c_{ilK-1}, d_{il-1}, d_{il})\right.$$
$$+ \left.\left.\sum_{k=1}^{K-1} \lambda_d p(1 - b_{ilk+1}, c_{ilk-1}, c_{ilk})\right)\right), \quad (16)$$

where $D \in \{0,1\}^{M \times (L-2)}$, and $\lambda_d$ is a positive real parameter.

This objective function has the quadratic and linear terms, which consist of the binary variables, $B$, $C$ and $D$. The proposed $l_0$-SVD is obtained by extending the optimization of (16) with the SVD preprocess in the $l_1$-SVD. To evaluate this method, we extract parameters $\{J, h\}$ from the observations $X$ and reconstruct $Z$ from obtained $B$ with (13).

## III. EXPERIMENT WITH THE QUANTUM ANNEALER

### A. Experimental Setup

All the quantum annealing process was performed on D-Wave Advantage Systems 4.1. The program codes for the proposed method were executed remotely through the cloud interface. We used a Python library, dwave-ocean-sdk 3.1.0, installed on a Windows 10 machine with an Intel Core i7-CPU and 64 GB of RAM. For the comparison of these results, simulated annealing computation was performed using dwave-neal as a reference for the same QUBO problems. We also used pyqubo 1.0.0 to extract $\{J, h\}$ from the program codes. For LASSO, OMP, SVD, and group-LASSO, we utilized publicly available libraries, including sklearn, numpy, and group-lasso.

In the evaluation, the original sparse vectors were sampled with random values for randomly selected nonzero entries. The observation vectors were simulated with the sampled sparse vectors and given observation matrices explained later. The proposed and the reference methods reconstructed the original sparse representations from the given observation, and the accuracy on the nonzero position detection was evaluated for each method. Specifically, the method that provided the reconstruction whose nonzero positions matched exactly for the original representations were given the score 1, and 0 otherwise. We repeated the evaluation procedure typically 100 times and used the average score as reconstruction success rates. In addition, we monitored the achieved values of (6) for the original representation $z$, quantized representation $q$, and reconstructed representation $s$ and those of (11) for $s$.

In the DOA estimation, since the observation matrix $A \in \mathbb{C}^{N \times M}$ is composed of the steering vectors, $a_{nm}$, the complex-valued elements at $(n, m)$ is expressed as

$$a_{nm} = \exp\left(2\pi i \frac{nd}{\lambda} \sin(\phi_m)\right), \quad (17)$$

where $n = 1,2 \dots N$ is the $n$-th sensor, $m = 1,2 \dots M$ is the $m$-th azimuth grid, $\lambda$ is the wavelength of the signal and $d$ is the distance of each array elements, typically, $d = 0.5\lambda$ is chosen.

Here, we use $\phi_m = \arcsin(2m/M - 1), m = 0,1 \dots M - 1$, instead of the monospaced grid and obtain,

$$a_{nm} = (-1)^n \exp\left(2n\pi i \frac{m}{M}\right). \quad (18)$$

This expression is similar to the top $N$ rows of the $M \times M$ discrete Fourier transformation, except that its sign depends on row indices. Although this grid arrangement becomes non-uniform, the similarity to the discrete Fourier transformation is useful. In addition, with this configuration, (18) provides a higher resolution than in (17) in the middle of the azimuth angles. For instance, when $M = 160$, it is calculated that the resolution inside angle $\pm 45°$ becomes less than 1°, which is finer than in the uniform spacing with $M=180$.

## B. Results for Single Observation

We observed the experimental results for an observation matrix $A$ defined by (18) with a size of $M$=32, 64, 128, 256 and 512, and an observation density of $N/M$=1/8, 1/4, 3/8 and 1/2. Here, we show some of the results in the combinations of the parameters $(N, M)$. For the original sparse vectors, nonzero values are sampled from a uniformed distribution on [0, 1]. We chose $K$=4 for the bit length of quantization, $\gamma_0$=0.001 for the $l_0$-regularization parameter, and $\lambda_c$=1.5 for the penalty parameter. $\gamma_0$ and $\lambda_c$ are supposed to be $0.5^d$ and $0.5d$ ($d = 0, 1 ...$), respectively. As the reference methods, we used LASSO and OMP from sklearn.

Fig. 1 shows the reconstruction success rates (left) and the loss values defined above (right) for $M$=32 and $N/M$=1/4. Similarly, for LASSO, we chose $\gamma_1$=0.0005, which gives the best results for $M$=32. For the nonzero detection, a threshold of 0.02 is used to avoid the false positive. In the left graph, the proposed method shows better success rates over other methods except for leftmost 2 plots. It was also observed for $N/M$=1/8 and 3/8, but not for $N/M$=1/2 and $M$>64. The degradation in the two leftmost plots was mainly caused by rounding small values to zero. The right graph shows that $L(s)$ does not reach $L(z)$, where $s$ and $z$ are the recovered vector and the original vector, respectively. From the comparison of $L(s)$ with $L(q)$, where $q$ is the quantized original vector, the quantization error was dominant in $L(s) - L(q)$ up to the leftmost 4 plots. From the match of $L(s)$ and $Lp(s)$ (the loss value in (11)), the constraints in (9) were satisfied for all plots.

Fig. 2 shows the results for $M$=32 and $N/M$=1/4 with 8-bit resolution ($K$=8). The left graph shows the degradation in the proposed method. This degradation gradually increased when $K$ was greater than 4. The right graph shows the overlapping plots of $L(q)$ and $L(z)$, which is reasonable for the increased $K$, and increase in $L(s) - L(q)$, which indicates increase in the insufficiency of the optimization.

Fig. 3 shows the results for $M$=512 and $N/M$=1/4 with 4-bit resolution ($K$=4). We chose $\gamma_1$=0.00006 for the LASSO, as a best value for $M$=512. In the left graph, the proposed method did not achieve the success rates for the reference methods. However the degradation of the proposed methods is not as large as in Fig. 2. The right graph shows that $L(q)$ gets close to $L(s)$, which is reasonable when $M$ increases. It also shows increase in $L(s) - L(q)$, which indicates that the optimization was insufficient, but not as large as that in Fig.2.

Fig. 4 shows the result for $M$=512, $N/M$=1/4 and $K$=4. In this experiment, Gaussian noise with standard deviation 0.01 were added to each observations. Similarly $\gamma_1$=0.00006 is chosen for the LASSO, and a threshold of 0.02 is chosen for the nonzero detection. The left graph shows degradation in both the proposed method and the reference methods compared with Fig.3. However, the degradation in the proposed method is not as large as that in the other methods. The right graph shows increase in $L(q)$ and $L(z)$ due to the observation noise. However, the increase in $L(q)$ is not observed in $L(s)$ except for the leftmost 3 plots, which implies that the proposed method is robust to the observation noise. Similar tendency were observed for the other $N$ and $M$. Especially, for $M$=32 and 64, the proposed method shows better success rates than the reference methods, which indicates that the proposed method is relatively robust to the observation noise.

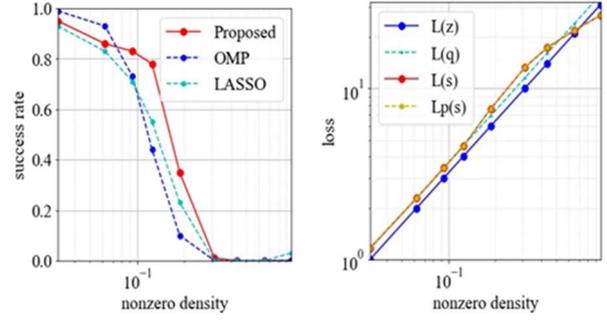

Fig. 1. Success rate of reconstruction (left) and loss values (right). The $x$-axis represents the density of nonzero representations. We evaluated 100 samples with a 32-dim representation and 8-dim observation with 4-bit quantization. Representation values are limited to [0, 1].

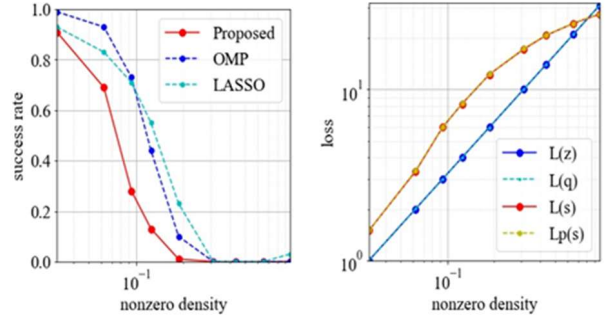

Fig. 2. Success rate (left) and loss values (right) for higher resolution. We sampled 100 problems with an 8-dim observation and 32-dim representation and evaluated our method with 8-bit quantization.

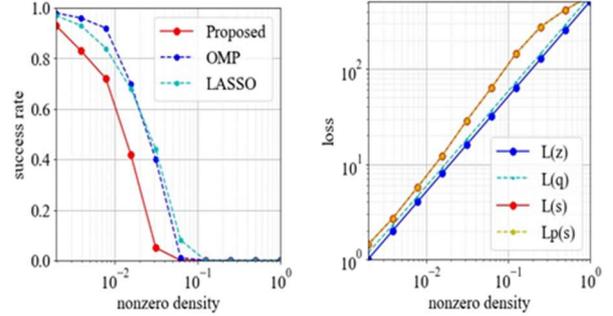

Fig. 3. Success rate (left) and loss values (right). We evaluated 100 samples with a 512-dim representation and 128-dim observation. We used our method with 4-bit quantization.

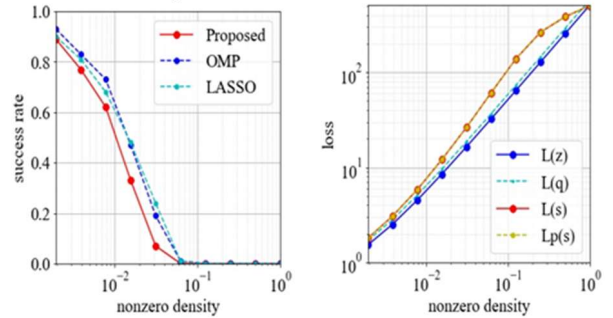

Fig. 4. Success rate (left) and loss values (right). We added white Gaussian noise (std: 0.01) for noisy observations. We evaluated 100 samples with a 512-dim representation and 128-dim observation. We used our method with 4-bit quantization.

## C. Results for Multiple Observations

In this section, we explain the results for the comparison between the proposed $l_0$-SVD with the conventional $l_1$-SVD. In this comparison, we observed the experimental results for the matrix $A$ defined by (18) with a size of $M$=32, 64 and 160, and an observation density of $N/M$=1/8, 1/4, 3/8 and 1/2. Here, we show some of the results in the combinations of these parameters.

For the original sparse vector $\zeta$, the nonzero entries were sampled from an uniformed distribution on [0, 1]. We duplicated $\zeta$ 16 times and added random values for each to generate 16 different sparse vectors. The additional random values were sampled from a normal distribution $N(0, \rho\zeta)$ with $\rho$ =0.1. Then, 16 observation vectors $x_1 \ldots x_{16}$ are generated with these sparse vectors and the observation matrix $A$. Following to the $l_1$-SVD, the preprocess with SVD provided the rectangular matrix $U \in \mathbb{C}^{N \times N}$ in the decomposed form $U\Sigma V^T$ of matrix $X = (x_1 \ldots x_{16})$. We used the leftmost 2 columns in $U$ as $X$ in (14) and obtained the solution $S$ from the proposed $l_0$-SVD. In the evaluation phase, we compared only the leftmost column $s_0$ with the original sparse signal $\zeta$. When the nonzero position of $s_0$ matched exactly that of $\zeta$, we judged the reconstruction was successfully completed. We chose $K$ =4 for the bit length of quantization, $\gamma_0$=0.0005 for the $l_0$-regularization parameter, and $\lambda_c = \lambda_d$ =1.5 for the penalty parameter. $\gamma_0$, $\lambda_c$ and $\lambda_d$ are supposed to be $0.5^d$ and $0.5d$ ($d = 0,1 \ldots$), respectively.

As a reference method, we used the $l_1$-SVD and the evaluation process for the $l_1$-SVD was the same as that for the $l_0$-SVD. In addition, we used the LASSO for the averaged observation as another reference method. We chose $\gamma_1$=0.00006 for both the $l_1$-SVD and the extended LASSO.

Fig. 5 shows the result for $M$=160, $N/M$ =1/4 and $K$ =4. The left and right graphs show the results for the clean ($\sigma$=0) and the noisy ($\sigma$=0.05) observations, respectively. In both the left and right graphs, the proposed method did not demonstrates degradation from the success rate of the reference methods. In addition, the proposed method were not robust than the conventional methods from the right graph. The results for the other observation densities were similar to this result. We also observed the results for $M$=64. In this case, the difference between the proposed and the reference methods was smaller than the results for $M$=160.

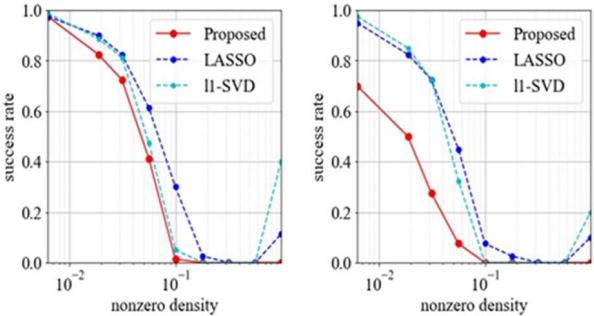

Fig. 5. Success rate for the clean (left) and the noisy (right) observations. We evaluated 50 samples of 16 shots, 40-dim observations generated from 160-dim representations. In the right graph, we added Gaussian noise (std: 0.05) for the observations.

IV. DISCUSSION

The proposed method showed some advantages than the conventional methods for small and coarse representations and noisy observations, but did not for other cases including multiple observations. In the following, we discuss the mechanism of the advantage and the cause of the degradation.

A. Effect of Quantization

To explain the improvement in several cases, we analyze a compensation effect in the discretized representation space. As in the experiment section, we analyze the value $\delta = L_0(s) - L_0(z)$, where $L_0$ is the objective function in (1), and $z, s$ are the original and the recovered vector, respectively. For the success reconstruction, the unique solution $s = z$ is required for the minimization of $L_0$, which corresponds to $\delta > 0, (\forall s \neq z)$ for $s, z \in \mathbb{R}^M$. When the observation with noise $n$ to be $x = Az + n$, $\delta$ is expressed as:

$$\delta = \frac{1}{2\gamma_0}\|Ae\|_2^2 + (\|e + z\|_0 - \|z\|_0) - \frac{1}{\gamma_0}e^T An, \quad (19)$$

where $e = s - z$. To obtain larger $P(\delta > 0)$ for the more stable reconstruction, the smaller variance from the last term in (19) and the larger positive bias from the other terms are required. When the quantization is applied to $z$ and $s$, $n$ in (19) is replaced with $n + Ar$, where $r$ is the quantization error in $z$. As a result, the quantization of z increases the variance in proportion to $\Delta^2/\gamma_0$, where $\Delta = 2^{-K}$ is the quantization width in $z$. On the other hand, the quantization of $s$ provides a positive finite value proportional to $\Delta^2/\gamma_0$ for the bias. This works as a margin to maintain the solution $s = z$. The ratio of the deviation and bias is roughly estimated as $\Delta/\sqrt{\gamma_0}$, thus, the coarser approximation of $z$ with larger $\Delta$ provides the more stable performance as in the experiment. Similarly, the smaller $\gamma_0$ can provide the more margin to improve the performance. However, we could not get it in the experiment. We consider the reason is insufficient resolution in the embedding of $\{J, h\}$.

B. Dependency on the problem size

As shown in the graphs for the loss values analysis, the optimizations did not reach the minimum for the case of the large sparse vector and the fine quantization. Theoretically, the quantum annealing can provide the global optimum in high probability with sufficient annealing time [7]. However the current quantum annealer has several known issues that prevent the ideal optimization. One of the issues is insufficiency in the annealing time, which is caused by the collapse of quantum states (coherence). It is estimated that the required time increases by exponential of number of qubits for obtaining the optimum with high probability. Thus, to avoid the insufficient optimization, the proposed methods needs to reduce the number of the binary variables. Another issue is the insufficient resolution that causes by the embedding of the coefficients $\{J, h\}$ on the analog values of the hardware. From the loss function (6), to obtain the difference in $l_0$-norm the required resolution is roughly estimated as $\sim \gamma_0$. However the resolution is limited by the he hardware. We consider this was the reason why the small $\gamma_0$ did not improve the performance.

V. CONCLUSION

The proposed $l_0$-regularized linear regression based on the QUBO formulation was evaluated for the performance on the sparse signal reconstruction with a commercially available quantum annealer. The method demonstrated the recovery success rates nearly equivalent to those of the conventional methods under the several conditions. The analysis of the results provided a hypothesis to the mechanism of the gain produced by the quantization. These results indicated the potential ability of the proposed method with the future improvement in the current quantum annealer.

REFERENCES

[1] R. Tibshirani, "Regression Shrinkage and Selection via the Lasso," J. R. Stat. Soc. Series B, vol. 58, pp. 267–288, 1996.

[2] D. L. Donoho, "Compressed sensing," IEEE Trans. Inf. Theory, vol. 52, pp. 1289–1306, April, 2006.

[3] D. Malioutov, M. Cetin, and A. S. Willsky, "A sparse signal reconstruction perspective for source localization with sensor arrays," IEEE Trans. Signal Process., vol. 53, pp. 3010–3022, August. 2005.

[4] E. J. Candès, "The restricted isometry property and its implications for compressed sensing," C. R. Math., vol. 346, pp. 589–592, May. 2008.

[5] G. Mallat and Z. Zhang, "Matching pursuits with time-frequency dictionaries," IEEE Trans. Signal Process., vol. 41, pp. 3397–3415, December. 1993.

[6] T. Kadowaki and H. Nishimori, "Quantum Annealing in the Transverse Ising Model," Phys. Rev. E, vol. 58, pp. 5355–5363, November. 1998.

[7] S. Morita and H. Nishimori, "Mathematical foundation of quantum annealing," J. Math. Phys., vol. 49, 125210, 2008.

[8] M. W. Johnson et al., "Quantum annealing with manufactured spins," Nature, vol. 473, pp. 194–198, May. 2011.

[9] https://pyqubo.readthedocs.io/en/latest/

[10] A. Lucas, "Ising formulations of many NP problems," Front. Phys., vol. 2, a5, February. 2014.

[11] M. Ohzeki, A. Miki, M. J. Miyama, and M. Terabe, "Control of automated guided vehicles without collision by quantum annealer and digital devices," Front. Comput. Sci. 1-9 November 2019.

[12] N. Ide, T. Asayama, H. Ueno, and M. Ohzeki, "Maximum Likelihood Channel Decoding with Quantum Annealing Machine," Proc. 2020 Int. Symp. Inf. Theory Appl., ISITA 2020, pp. 91–95, 2020.

[13] R. Ayanzadeh, S. Mousavi, M. Halem, and T. Finin, "Quantum annealing based binary compressive sensing with matrix uncertainty," https://arxiv.org/abs/1901.00088.

[14] R. Ayanzadeh, H. Milton, and T. Finin, "An ensemble approach for compressive sensing with quantum annealers," IEEE International Geoscience and Remote Sensing Symposium, pp. 3517–3520R, 2020.

[15] H. Friedman and J. Tibshirani, "Regularization Path for Generalized Linear Models by Coordinate Descent," J. Stat. Softw., vol. 33, pp. 1–22, February. 2010.

[16] R. Zibulevsky and M. Elad, Efficient Implementation of the K-SVD Algorithm using Batch Orthogonal Matching Pursuit Technical Report, CS Technion, April. 2008.